\documentclass[twocolumn,superscriptaddress]{revtex4}
\usepackage{amsmath,amssymb}
\usepackage[hiresbb,final]{graphicx}

\newcommand{\celsius}{\textrm{C}}
\newcommand{\pascal}{\textrm{Pa}}
\newcommand{\poise}{\textrm{P}}
\newcommand{\centi}{\textrm{c}}
\newcommand{\milli}{\textrm{m}}
\newcommand{\micro}{\mu}
\newcommand{\nano}{\textrm{n}}
\newcommand{\meter}{\textrm{m}}
\newcommand{\liter}{\textrm{l}}
\newcommand{\second}{\textrm{s}}
\newcommand{\tot}{\textrm{tot}}
\newcommand{\short}{\textit{s}}
\newcommand{\lon}{\ell}

\begin{document}
\title{Extracting the hydrodynamic resistance of droplets from their behavior in microchannel networks}
\author{Vincent Labrot}
\affiliation{Rhodia/CNRS Laboratory of the Future, 178 avenue du Docteur Schweitzer, 33608, Pessac, France}
\author{Michael Schindler}
\affiliation{Laboratoire Physico-Chimie Th\'eorique, UMR ``Gulliver'' CNRS-ESPCI 7083, 10 rue Vauquelin, 75231 Paris cedex 05}
\author{Pierre Guillot}
\author{Annie Colin}
\author{Mathieu Joanicot}
\affiliation{Rhodia/CNRS Laboratory of the Future, 178 avenue du Docteur Schweitzer, 33608, Pessac, France}

\begin{abstract}
The overall traffic of droplets in a network of microfluidic channels is
strongly influenced by the liquid properties of the moving droplets. In
particular, the effective hydrodynamic resistance of individual droplets plays a
key role in their global behavior. We here propose two simple and low-cost
experimental methods for measuring this parameter by analyzing the dynamics of a
regular sequence of droplets injected into an ``asymmetric loop'' network. The
choice of a droplet taking either route through the loop is influenced by the
presence of previous droplets which modulate the hydrodynamic resistance of the
branches they are sitting in. We propose to extract the effective resistance of
a droplet from easily observable time series, namely from the choices the
droplets make at junctions and from the inter-droplet distances. This becomes
possible when utilizing a recently proposed theoretical model, based on a number
of simplifying assumptions. We here present several sets of measurements of the
hydrodynamic resistance of droplets, expressed in terms of a ``resistance
length''. The aim is twofold, (1)~to reveal its dependence on a number of
parameters, such as the viscosity, the volume of droplets, their velocity as well as the
spacing between them. At the same time~(2), by using a standard measurement
technique, we compare the limitations of the proposed methods. As an important
result of this comparison we obtain the range of validity of the simplifying
assumptions made in the theoretical model.
\end{abstract}

\maketitle

\section{Introduction}

Droplets in microfluidic channels are widely used as microreactors for chemical
analysis, for chemical kinetics studies, or, for example, as a tool for
continuous nanoparticle
synthesis~\cite{JoaAjd05,CristobalETAL06,TiceETAL03,DenTsoHatDoy05,ZheRoaIsm03}.
Propelled by the necessity to accommodate parallel processes and to sample or
sort droplet populations, the demands on microfluidic networks to implement
certain procedural targets keep on increasing. One strategy towards the parallel
processing of many droplets are \emph{self-regulated} channel networks in which
it is not necessary to control the route of individual droplets by external
means (valves) but in which a well-chosen channel design leads to a autonomous
behavior of the droplet
train~\cite{CristobalETAL06,PraGer07,ThorsenETAL01,SonTicIsm03}. The advantage
of this approach is evident as the channel geometries stay simple, and less
external devices are required. A drawback of the approach is an increased
complexity in the understanding of the droplet dynamics, making it more
difficult to find robust implementations of given
tasks~\cite{WilBarKloMaiTab06,GarFueWhi05,SchAjd08}. General features of
self-regulated droplet dynamics are the global coupling of the dynamics, meaning
that nearly all droplets influence each other, and the strong nonlinearity in
the equations describing the droplet positions as a function of time.

The non-local and non-linear droplet dynamics has been described in
Ref.~\cite{SchAjd08} by a simplified theoretical model. In this model, each
droplet, sitting in one connecting channel, increases the hydrodynamic
resistance of this channel by a given constant amount. By its presence it
changes the flowrate in the channel, and in turn it changes also the flowrates
in possibly all other channels of the network. When a droplet arrives at a
junction, a rule is required to decide which of the possible exits is to be
taken. It will here be assumed that it takes the channel with the highest
instantaneous flowrate. The simplified model of Ref.~\cite{SchAjd08} was able to
address several intriguing properties of different network layouts, such as the
reversibility of the droplet dynamics and the influence of network symmetry. The
model was proposed as a simple tool for finding robust dynamical behavior and to
quantify its response to changes in the driving parameters and the geometrical
parameters of the network. While applying the model calculations to existing
devices, it turned out that the most important parameter for robust behavior is
the \emph{effective hydrodynamic resistance of droplets}. Compared to its
important role in the overall dynamics of droplets in a network, this property
has not attracted sufficient interest, at least for droplets of a viscous fluid.
As a first guess, it was therefore assumed to be a constant in the model.

In this paper, we present results from three methods for measuring the
hydrodynamic resistance of droplets being transported in microchannels: A
standard method that employs a pressure sensor as a reference, and two new
self-regulated low-cost methods which do not require laborious external
equipment. These novel methods follow the idea of the abovementioned
self-regulated networks. We here propose to \emph{make use of the overall
droplet dynamics}, which is easily observable with a video device, in order to
\emph{extract effective properties such as the hydrodynamic resistance}. As we
make use of self-regulating droplet dynamics, which requires a theoretical model
for its interpretation, the target here is twofold: \emph{As long as the
assumptions of the model from Ref.~\cite{SchAjd08} hold, the property extracted
from the measurement is really the hydrodynamic resistance of the droplets. At
the same time we verify these very assumptions by checking the consistency of
the outcome and by directly measuring the hydrodynamic resistance of droplets
with the standard pressure sensor.}

The paper is organized as follows: First, in Sec.~\ref{sec:principles} we explain the
principles of the methods the specific advantages and limitations of each of
them: First, the well established method based on the use of an external
pressure sensor, and then, the principle of the new methods relying on the flow
of a regular train of droplets in a ``loop'' device. We distinguish between two
different sizes of the loop, giving different weights to the underlying
assumptions on which the interpretation is based. In Sec.~\ref{sec:details} we
describe the experimental procedure we used for creating the device.
Section~\ref{sec:results} contains the experimental findings of the two
variants of the loop method and of the method using a pressure sensor. There, we
discuss in detail the implication of the measurements, being in favor of the
simple model or putting its assumptions into question.

\section{Principles of the methods}
\label{sec:principles}
\begin{figure}[tb]%
  \centering
  \includegraphics[width=\linewidth]{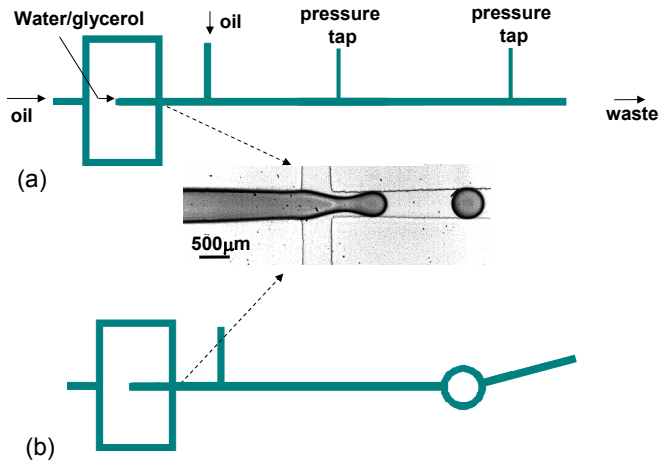}%
  \caption{Schematic pictures of the devices: (a)~A straight channel with two
  pressure taps for the method using the pressure sensor, and (b)~an asymmetric
  circular loop. Droplets are generated where the first oil and water/glycerol
  inlets meet. The third inlet allows to adjust the distance between the
  droplets. Channels and inlets have a rectangular cross-section,
  $500\,\micro\meter$ wide and $300\,\micro\meter$ high. Only the two channels
  to the pressure sensors are narrower ($100\,\micro\meter$ wide).}%
  \label{fig:schema}
\end{figure}%

\subsection{The pressure sensor method}

Experiments were conducted using a differential pressure sensor as described by
by Adzima \textit{et al.}~\cite{AdzVel06}. At a distance of $127\,\milli\meter$,
two pressure taps were connected to a long straight channel, as depicted in
Fig.~\ref{fig:schema}a. At each of them the pressure is measured by a pressure
sensor (Validyne~P55D), equipped with a $1.25\,$psi-membrane. By comparing the
pressure differences in two situations, namely with and without droplets---while
keeping the total flowrate~$Q$ the same---we infer the additional pressure
caused by the presence of $N$~droplets between the taps. The hydrodynamic
resistance~$R_d$ of each droplet is then given by
\begin{equation}
  R_d = \frac{\Delta P^\ast - \Delta P}{N\,Q},
\end{equation}
where $\Delta P^\ast$ and $\Delta P$ are the pressure differences between the
taps, with and without droplets respectively~\cite{AdzVel06}. In the discussion
below it will prove convenient to express hydrodynamic resistances in terms of
lengths. This is natural for channels without droplets where the hydrodynamic
resistance~$\bar R_i$ of a channel~$i$ is known to scale linearly with its
length~$L_i$,
\begin{gather}
  \label{res_length}
  L_i = \frac{w h^3}{A\eta_\Phi}\,\bar R_i \\
  \text{with}\quad A := 12\Bigl(1 - \frac{192w}{\pi^5h}\tanh\frac{\pi h}{2w}\Bigr)^{-1}.
\end{gather}
Here, $w$~and $h$ are the channel width and height, respectively; $\eta_\Phi$~is the
viscosity of the continuous phase; and the dimensionless parameter~$A$ depends
on the shape of the channel cross-section, which is rectangular
here~\cite{MorOkkBru05}. We apply the same factor of linearity from
Eq.~\eqref{res_length} also to the effective resistances~$R_d$ of the droplets
in order to define their resistance length~$L_d$,
\begin{equation}
  L_d := \frac{w h^3}{A\eta_\Phi}\,R_d.
\end{equation}%
\begin{figure}[tb]%
  \centering
  \includegraphics{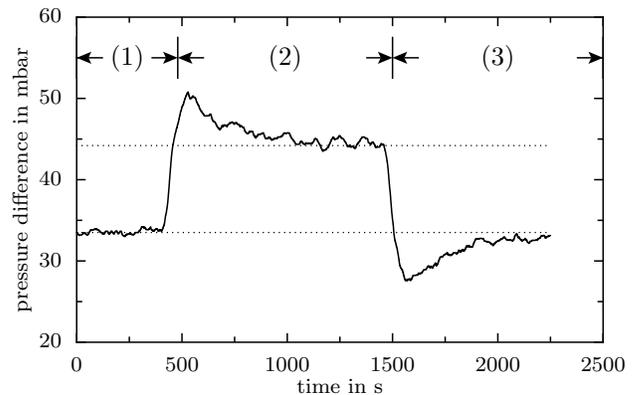}%
  \caption{Time series of the pressure difference along a straight channel, as
  measured with the pressure sensor. The droplet input has been changed
  subsequently: First~(1) without droplets, then~(2) with them, and~(3) last
  without droplets. The total flow rate has been kept constant. The dotted lines
  are guides for the eye. Flow rates of the oil and the water inlets in the
  droplet generator are respectively $8\,\micro\liter/\min$ and
  $7\,\micro\liter/\min$. The second oil inlet allows to adjust the distance and the
  velocity of droplets, here $70\,\micro\liter/\min$. The parameters of the
  droplets are~$\eta_d=0.001\,\pascal\,\second$, $\Omega_d=85\,\nano\liter$,
  $v_\tot = 7.5\,\milli\meter/\second$, $\lambda = 6\,\milli\meter$.}%
  \label{pressure_timeseries}
\end{figure}%

The pressure sensor is used below in Sec.~\ref{sec:results} to compare the
results of the alternative passive methods presented in the following
subsections. It is important to note that this equipment is quite costly
(several hundreds of dollars) and that it presents a number of limitations in
terms of response time and ease of use: The smaller the probed channel, the
larger is the equilibration time. In the case of our channels, which are quite
large ($300\,\micro\meter$ high and $500\,\micro\meter$ wide), this
equilibration time was found to be of the order of ten minutes.
Figure~\ref{pressure_timeseries} shows the pressure difference at the two taps
as a function of time, exhibiting the slow relaxation of the signal at the
beginning and at the ending of a droplet series. The following passive methods
allow to get rid of these constraints.

\subsection{The loop devices}
\begin{figure}[tb]%
  \centering
  \includegraphics{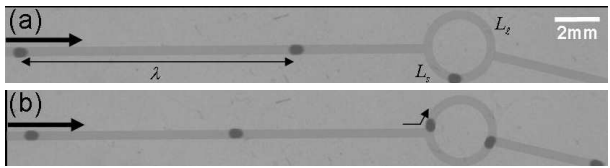}%
  \caption{Snapshots of the small-loop device, fed by a regular sequence of
  droplets: (a)~all droplets take the short route, the incoming distance is
  large, $\lambda>\lambda_\text{max}$. (b)~alternating droplet decisions due to
  interactions between droplets, $\lambda<\lambda_\text{max}$. The arrows
  indicate the flow direction. Parameters are $L_s=4.25\,\milli\meter$,
  $L_\lon/L_\short=1.18$.}%
  \label{fig:small_loop}%
\end{figure}%
\begin{figure}[tb]%
  \centering
  \includegraphics{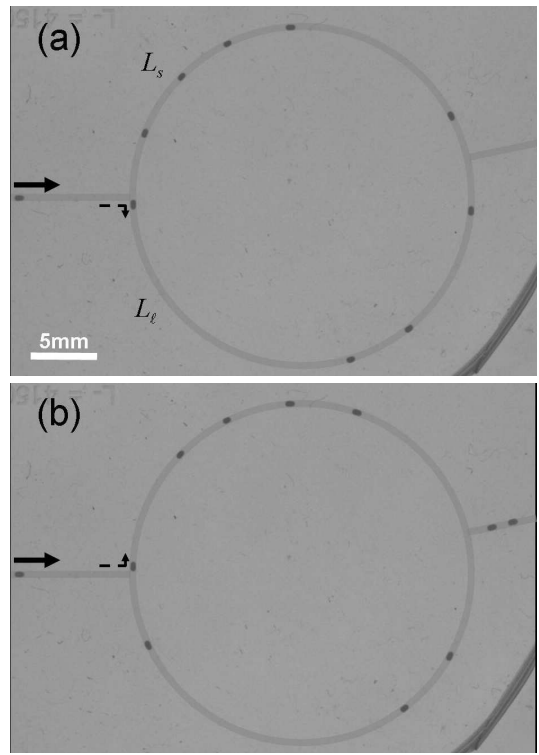}%
  \caption{Snapshots of the large-loop device: (a)~the arriving droplet chooses
  the long branch, corresponding to the relation~\eqref{eq:ll_prelim}, here
  $\bar R_\lon + 3R_d < \bar R_\short + 5R_d$. (b)~the next arriving droplet
  takes the short route, $\bar R_\short + 4R_d < \bar R_\lon + 3R_d$. The dotted
  arrows indicate the direction chosen by the droplets. Parameters are $L_\short
  = 37\,\milli\meter$, $L_\lon/L_\short = 1.16$.}%
  \label{fig:large_loop}%
\end{figure}%

We here propose two variants of a new method which allows to extract the
hydrodynamic resistance of droplets from the observed behavior of regular
droplet trains. This approach has the advantage over the standard pressure
sensor that it is cheap and does not require taps which pose an additional
source of perturbation of the flow. Furthermore, it can be applied without
problems to very small channels, where the pressure sensor technique becomes
inapplicable. Of course, our method has its limitations: The main limitations
come from the hypotheses made in the theoretical model on which the
interpretation of recorded droplet behavior is based. It will turn out in
Sec.~\ref{sec:results} that these assumptions are not always satisfied.
Nevertheless, in those regions where they do hold, the new method presents a
real alternative to the pressure sensor.

We use a very simple network of microfluidic channels as depicted in
Fig.~\ref{fig:schema}b. Disregarding the part in which droplets are generated,
it consists of two long channels with an asymmetric circular loop in-between. A
periodic sequence of droplets is generated and injected into this loop from one
side. The droplets then take either the short or the long branch of the loop,
depending on how many droplets the respective branches contain: For the first
droplet, the shorter branch is certainly favorable. It has a higher flowrate and
a smaller hydrodynamic resistance. The arriving first droplet then increases the
hydrodynamic resistance of this shorter branch, possibly so much that the
following droplet takes the longer route. Upon continuation, a complicated
sequence of droplet ``decisions'' is generated, exhibiting complex periodicity.
The same network topology has been used in Refs.~\cite{GarFueWhi05,SchAjd08} to
investigate the reversibility of the droplet dynamics. The droplet traffic in
the network was recorded by a charge-coupled device (CCD)~camera and processed with a homemade Matlab
program. Such, we extracted from the movies the velocities of all droplets, the
number of droplets in each channel, and the temporal and spatial periods of the
incoming droplet sequence. These properties were then plugged into the
theoretical model in order to extract the hydrodynamic resistances, according to
the two methods described in the following:

The flow through the devices is governed by flow equations which resemble
Kirchhoff's law for electric circuits. At each junction, the sum of incoming and
outgoing flowrates must vanish. This means for the loop device
\begin{equation}
  Q_\tot = Q_\short + Q_\lon,
\end{equation}
with~$Q_\tot$ the total flowrate in the input/output channels, and
with~$Q_\short$ and~$Q_\lon$ the flowrates in the short and long branches of the
loop. Here and in the following, the indices $\short$~and $\lon$ stand for the
short and the long channel, respectively. The second set of equations provides
the relationship between the pressure difference along a channel and the flowrate
passing through it,
\begin{equation}
  \label{eq:defin_R}
  (\Delta P)_i = R_i\,Q_i.
\end{equation}
At the moment, this equation is nothing but the definition of the total
hydrodynamic resistance~$R_i$ of channel~$i$. Without any droplets, it is known
to provide a linear relation, in that the resistance does not depend on the
total flowrate.

Before describing the dynamics of droplets in the two geometries we used, let us
summarize the assumptions which led to the simplified model and which will be
investigated experimentally in the following sections:
\begin{enumerate}
\item[(A)] All droplets in a channel~$i$ have the same velocity
  velocities~$v_i$. Its value is proportional to the total flowrate~$Q_i$ in the
  channel (oil + water/glycerol), with a constant factor of proportionality,
  \begin{equation}
    v_i = \frac{\beta}{S}\,Q_i.
  \end{equation}
  Here, $S$~is the cross-section which does not change. $\beta$~was found to be
  approximately $1.6$ in our experiments.
\item[(B)] Each droplet adds the same amount~$R_d$ to the total resistance~$R_i$ of a
  channel~$i$, such that
  \begin{equation}
    R_i = \bar R_i + n_i R_d,
  \end{equation}
  with $n_i$~the number of droplets in the channel, and with~$\bar R_i$
  its resistance in the absence of droplets.
\item[(C)] The resistance~$R_d$ of droplets should be a constant. In particular,
  $R_d$~should not depend on the flowrate~$Q_i$ (nor on~$v_i$).
\item[(D)] At junctions, droplets take the route with highest flowrate leading
  away from the junction.
\end{enumerate}
\begin{figure}[tb]%
  \centering
  \includegraphics{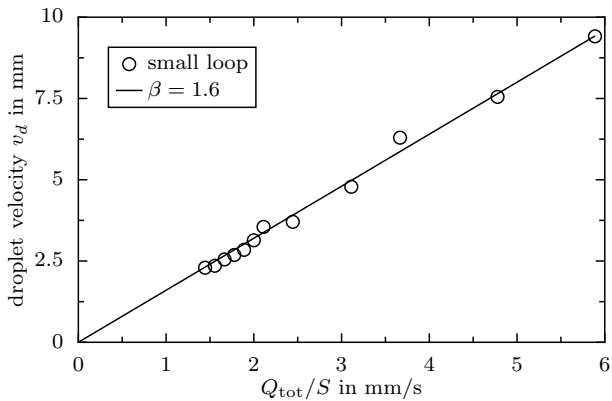}%
  \caption{Experimental test for the linear relation between droplet
  velocity~$v_i$ and the flowrate~$Q_i$ (Assumption A). The data were
  acquired in a straight channel. The parameters
  are~$\eta_d=0.76\,\pascal\,\second$,
  $\eta_\Phi = 0.02\,\pascal\,\second$,
  $\lambda = 1.7\text{ to }2.8\,\milli\meter$,
  $\Omega_d=55\,\nano\liter$.}%
  \label{vel_vs_qtot}
\end{figure}%

Some comments on the assumptions and their expected validity are in order at
this point: The linearity of velocity and total flowrate~(A) has been observed
under experimental conditions, both in Fig.~\ref{vel_vs_qtot} and in
Ref.~\cite{EnglETAL05}. Assumption~(B) will prove false if there is cooperative
behavior between the droplets. Such effects will not occur as long as the
droplets are sufficiently distant. We will quantify the necessary distance in
the experimental section. Assumption~(B) will also fail if the droplets have
different sizes or if they are smaller than the channel and are thus not forced
to be centered. Assumption~(C) is nothing else than the linearity of
Eq.~\eqref{eq:defin_R}. In fact, we will present deviations which are important
for small velocities, but not for larger ones. Also in case of long plug-like
droplets, a nonlinearity in Eq.~\eqref{eq:defin_R}, and therefore a
velocity-dependence of~$R_d$ must be expected. Such effects have been reported
for bubbles~\cite{Bretherton61,WonRadMor95b,HodJenRal04,LinkETAL04}. By
construction of our device, that is by virtue of its simple geometry, we do not
require assumption~(C) for the method to be valid. Finally, assumption~(D)
will not be questioned here. The precise criterion, whether it is the highest
flowrate, the highest pressure drop, or the largest velocity which counts, does
not seem to be important in our case as we use locally symmetric junctions and
the same channel cross-sections everywhere.

\subsubsection{The small-loop device}
\label{small}

The first variant of the proposed device comprises a small loop, as shown in
Fig.~\ref{fig:small_loop}. The two branches of the loop are approximately
ten~times longer than they are wide. The idea of the method is to send in a
periodic sequence of droplets and to vary slowly their initial distance. If one
starts with quite a large distance, the droplets will all take the short route
and leave the loop before the next one arrives, see Fig.~\ref{fig:small_loop}a.
For a continuously reduced initial distance~$\lambda$, one finds a sudden change in the
droplet behavior at a specific distance~$\lambda_\text{max}$. The droplets start
to influence each other: The following droplet will take the long route because
the previous one still enhances the resistance of the short branch. This
situation is depicted in Fig.~\ref{fig:small_loop}b. Of course, in order to make
the method work, the (yet unknown) hydrodynamic resistance of a single droplet
must be larger than the difference in resistances of the channels when no
droplets populate them. The applicability can therefore be adjusted by choosing
the length ratio $L_\lon/L_\short$. The hydrodynamic resistance of the droplet is
determined from the value of the initial distance~$\lambda_\text{max}$, at which
the droplets stop taking the entire short branch, by applying the simple
Eq.~\eqref{eq:small_loop} below, which we will derive now.

The initial distance~$\lambda_\text{max}$ is obtained as the distance which an
approaching droplet can travel while the previous one traverses the short branch.
The time~$T$ which a droplet stays in the short branch is expressed as follows,
when assumption~(A) is used,
\begin{equation}
  T = \frac{L_\short}{v_\short} = \frac{SL_\short}{\beta Q_\short},
\end{equation}
where~$L_\short$, $v_\short$, and $Q_\short$~are the length, the velocity and
the flow rate in the short arm. In the same time span, another droplet in the
incoming channel covers the length
\begin{equation}
  \lambda_\text{max} := T \frac{\beta}{S}  Q_\tot,
\end{equation}
which leads to the elimination of the parameters $\beta$~and~$S$, and which leaves us with
\begin{equation}
  \frac{\lambda_\text{max}}{L_\short}
  = \frac{Q_\tot}{Q_\short}
  = 1 + \frac{Q_\lon}{Q_\short}.
\end{equation}
Now, we make use of
assumption~(B), together with Eq.~\eqref{eq:defin_R}. As the pressure
difference is the same in both branches of the loop, we obtain for the ratio of
their flowrates
\begin{equation}
  \frac{Q_\lon}{Q_\short}
  = \frac{R_\short}{R_\lon}
  = \frac{L_\short+L_d}{L_\lon}.
\end{equation}
Here, it proves convenient to express the hydrodynamic resistances in terms of
lengths, as defined in Eq.~\eqref{res_length}. Combining the last two equations, we
find the result already given in Ref.~\cite{SchAjd08},
\begin{equation}
  \label{eq:small_loop}
  \frac{\lambda_\text{max}}{L_\short}
  = 1 + \frac{L_\short}{L_\lon} + \frac{L_d}{L_\lon},
\end{equation}
which can easily be resolved for~$L_d$. Note that we did not make use of
assumption~(C) in this short derivation. Equation~\eqref{eq:small_loop}
therefore holds for any non-linear relationship between~$Q_\short$
and the resistance~$R_d(Q_\short)$ of the droplet. It is the simple geometry of
the device which allows to circumvent the simplifying assumption~(C), which is
not possible in extended networks. Note also the nice property of this formula
that the hydrodynamic resistance length of the droplet is only given by
geometrical properties, namely the two lengths of the branches, and by the
experimentally determined initial distance~$\lambda_\text{max}$ between
droplets. This means that once the device has been elaborated, the measure of
the droplet resistance is simply obtained from observing the droplets taking
different routes through the loop.

Note also that the determination of the droplet resistance from this device
relies strongly on the assumption~(A), and only weakly on the assumption~(D).
Assumption~(C) is not used at all, and assumption~(B) is unimportant, since we
treat one droplet per channel ($n=1$) which is the trivial case for~(B).

The limitation of the here described small-loop method is that the two branches
are quite short. A violation of the above derivation might occur, in that the
resistance~$R_d$ depends on the instantaneous distances between droplet and the
two junctions, both of which perturb the flow in a non-trivial manner, leading
to an effective hydrodynamic interaction between them. In this aspect, the
smaller the channel cross-section is, the better the small-loop device works.
This will lead to a more uniform flow and at the same time increase the
resistance of droplets.

\subsubsection{The large loop device}
\label{large}

We propose a second variant of the loop device which allows to determine the
hydrodynamic resistance of droplets with a different weighting of the
assumptions. This time, the loop is much larger compared to the previous
geometry, see Fig.~\ref{fig:large_loop}. Like for the small loop device, a
regular train of droplets is injected into the main channel. This time, there can
be several to many droplets in each branch of the loop, and their resistance is
determined by counting the droplets in each branch at every time a new droplet
has made its decision. Thus, the droplet dynamics depends on the model
assumptions in a different way: According to assumption~(D) each droplet
entering the loop systematically goes into the branch of lower hydrodynamic
resistance. And, since each droplet adds a certain resistance to the channel
(assumption~B), it tends to equilibrate the hydrodynamic resistances of the two
branches of the loop. The same principle has been previously used in the
analysis of the hydrodynamic resistance of bubbles~\cite{FuerstmanETAL07},
however only in the case of a symmetric loop and without the complementary
device with a small loop.

During the run of the experiment, the total hydrodynamic resistances of the two
branches change as functions of the numbers of droplets they already contain. We
thus have to track the relation between these two resistances,
\begin{equation}
  \label{eq:ll_prelim}
  \bar R_\short + n_\short R_d \gtrless
  \bar R_\lon + n_\lon R_d,
\end{equation}
where both comparison operators are found. Each time a droplet has chosen a
route, we gain some information about which branch of the loop has had the
smaller total resistance. In fact, each time a droplet chooses the longer
branch, we obtain a lower bound for~$R_d$; the short branch provides an upper
bound. When resolving Eq.~\eqref{eq:ll_prelim} for~$R_d$, rewritten in terms of
the corresponding length~$L_d$, we obtain these bounds as
\begin{equation}
  \label{eq:large_loop}
  \biggl[\frac{L_\lon-L_\short}{n_\short - n_\lon}\biggr]_{\to\textit{long}}
  < L_d <
  \biggl[\frac{L_\lon-L_\short}{n_\short - n_\lon}\biggr]_{\to\textit{short}}.
\end{equation}
The subscripts ``$\to$\textit{long}'' and ``$\to$\textit{short}'' indicate that
the specified bound applies only if the long/short branch has been chosen.
Observing many droplets will successively narrow the bounds on~$L_d$, leaving
finally an interval which depends on both the initial droplet spacing~$\lambda$
and on the length ratio~$L_\lon/L_\short$. A large loop which is nearly
symmetric leads to a good accuracy of the method. Figure~\ref{fig:large_loop}
shows two examples of how Eq.~\eqref{eq:large_loop} is applied, leading in
Fig.~\ref{fig:large_loop}a to a lower bound, and in Fig.~\ref{fig:large_loop}b
to an upper bound.

In terms of the assumptions required for the analysis, this large-loop method
appears to be complementary to the small-loop variant. We here make heavy use of
assumption~(B) in that we assume the droplets to add up their individual
resistances. On the contrary, the small-loop proposition was mainly based on
assumption~(A) which is not really required here, because the time the droplets
stay in in both branches of the loop are not used. Also the role of the droplet
velocity is different. While the flowrates in the two branches of the small-loop
device can be very different, they are automatically balanced here---at least
their average values, with fluctuations due to the fluctuating numbers of
droplets.

\section{Experimental details}
\label{sec:details}

We made the microfluidic devices out of poly(dimethyl-siloxane) (PDMS) using
standard photo-lithographic and soft-lithographic
techniques~\cite{QinXiaWhi96,DufMcDSchWhi98}. A transparency mask of
$20\,000\,\textrm{dpi}$ resolution (CAD/Art Services~Inc., California) was used
in 1:1~contact photolithography with SU-8~photoresist (MicroChem, Newton, MA) to
generate a negative ``master'' consisting of patterned SU-8 photoresist on a
silicon wafer with a flat surface. Positive replicas were formed by molding a
5:1 mixture of~PDMS with a temperature-active cross-linker against the master.
After curing the PDMS layer for 20~minutes at $65^\circ\celsius$, it was peeled
off the silicon wafer, and inlet and outlet holes were punched. A glass slide,
serving as the rigid substrate of the device, was covered by a thin layer of a
20:1 mixture of PDMS/cross-linker and cured during 40~minutes at
$65^\circ\celsius$. The PDMS surfaces were then activated with an UVO-cleaner
(Jelight 144AX)~\cite{HarrisonETAL04} for $40\,\second$, and both PDMS surfaces
were immediately brought together. An irreversible seal was formed between the
PDMS surface and the glass substrate.

The resulting rectangular microchannels are $300\,\micro\meter$ high and
$500\,\micro\meter$ wide. Droplets were fabricated from a droplet generator with
a cross-junction (Fig.~\ref{fig:schema}): Inlets for different liquids, one of
oil (viscosity $\eta_\Phi=20\,\centi\poise$) and one for a glycerol/water
mixture, lead into a channel located downstream the junction. A
colorant~(Bromothymol blue) was added to this aqueous phase in order to increase
the contrast of the droplet. External connections to syringes (Hamilton,
$2.5$--$10\,\milli\liter$) were made using polyethylene (PE20) medical tubing
(outer diameter: $1.09\,\milli\meter$). The syringes were contracted by three
programmable syringe pumps (Cetoni, Nemesys) controlled by a computer. An
additional oil inlet further downstream served to vary the spatial period
between droplets without modifying the volumes of the droplets. Their volumes
were inferred from all three incoming flow rates and the measured droplet
distance.

\section{Results and discussion}
\label{sec:results}

In this section we present several sets of measurements in which different
parameters are varied. We measure their influence on the hydrodynamic resistance
of the droplets, in order to see to what extent the assumptions (A) to~(D) may
be applied. The varied parameters are the volume~$\Omega_d$ of the droplets,
their velocity, their viscosity~$\eta_d$ and their initial distance~$\lambda$.
All these parameters influence the hydrodynamic resistance of droplets---which,
strictly speaking, invalidates assumptions~(B) and~(C). Nevertheless, we will
show that reliable results can be obtained with the newly proposed methods.
Their reliability is enforced by the comparison with the results from the
pressure sensor as a reference. The first two presented results coincide well
and show that the proposed small-loop and large-loop devices can be used to
determine the hydrodynamic resistance of droplets.

From the remaining results we learn that the assumptions underlying these
methods are not fulfilled for all driving parameters. On first sight, this
appears as bad news for the methods we propose. However, we also find that there
is always a parameter regime in which the assumptions are indeed valid.
Concretely, we find that both the droplet velocity and the inter-droplet
distance must be kept sufficiently large. If we ensure that the velocity does
not drop below a critical value, while varying for example the distance, the
result is perfectly valid. We may use such a situation to extract the
hydrodynamic resistance of the droplets. The results given in this section
therefore present both a direct verification of the assumptions of the employed
model and a measurement of the hydrodynamic resistance.

\subsection{Validity of the assumption~(A)}

Independently of the parameter variations, Fig.~\ref{vel_vs_qtot} shows that the
assumption~(A) holds in the performed experiments.

\subsection{Influence of the droplet volume}
\begin{figure}[tb]%
  \centering
  \includegraphics{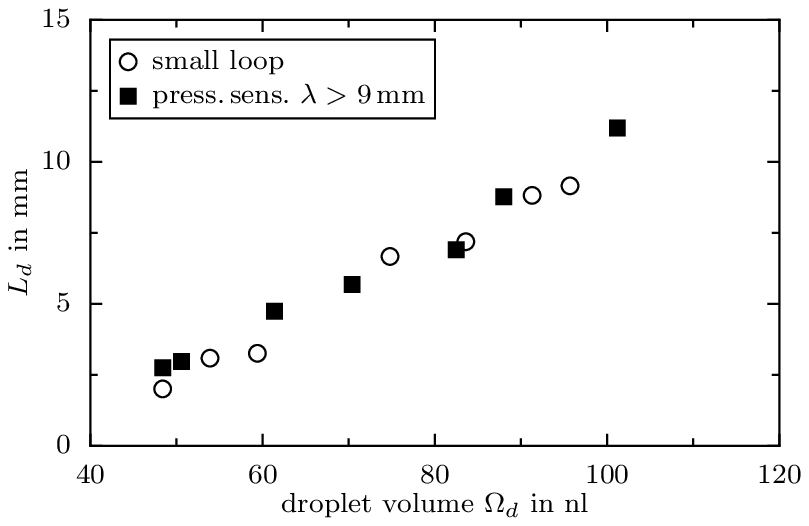}
  \caption{The resistance length~$L_d$ as a function of the droplet volume:
  measured with the small-loop device and with the pressure sensor. The
  parameters are~$\eta_d=1.4\,\pascal\,\second$,
  $\eta_\Phi=0.02\,\pascal\,\second$,
  $L_\short=4.25\,\milli\meter$,
  $L_\lon/L_\short=1.18$,
  $v_\tot\approx 6\,\milli\meter/\second$, and
  $\lambda>9\,\milli\meter$.}%
  \label{Ld_vs_vol}
  \centering
  \includegraphics{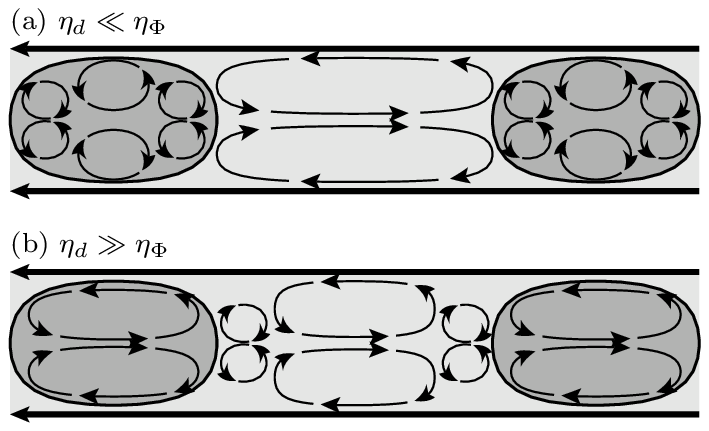}%
  \caption{The topology of the counter-rotating eddy pairs in front, behind and
  within the droplets moving in a channel. The velocity streamlines are those
  seen from a frame co-moving with the droplets: The channel walls move to the
  left. Depending on the viscosity ratio, between one and three eddy pairs can
  be found in the droplets. More eddies are possible, depending on the shape of
  the droplets, thus on the surface tension.}%
  \label{eddies}
\end{figure}%
The influence of the volume of droplets on their hydrodynamic resistance has
been investigated with the pressure sensor and with the small-loop device. The
results in Fig.~\ref{Ld_vs_vol} depict a strong increase of the resistance with
the volume, which is confirmed by the measurements of the pressure sensor. Both
methods yield coinciding values for the resistance. The increase can be
understood by the fact that with growing volume the droplets fill the channel
more closely and become longer. Similar effects have been reported in
Refs.~\cite{AdzVel06,EnglETAL05}. Droplets of $48\,\nano\liter$ are nearly
spherical when looked at from above. Larger droplets are more elongated, giving
rise to thin films of oil between the droplets and the channel walls.

The viscous dissipation in these thin films is one of three possible reason for
the enhancement of hydrodynamic resistance due to the presence of a droplet:
Generally speaking, each droplet perturbs the surrounding flow field, which
would otherwise be an axial Poiseuille-type flow, see Fig.~\ref{eddies} and
Ref.~\cite{MalschETAL08}. The kinematic boundary condition at the boundary of
the droplet enforces the presence of several counter-rotating eddies inside and
outside the droplet. Figure~\ref{eddies} contains only those eddies which
are necessary to ensure a continuous velocity field. In particular, the tangent
velocity has to be continuous at the droplet boundary. Due to their different
viscosities, the eddies inside and outside the droplet may contribute
differently to the total hydrodynamic resistance of a droplet. A third
contribution comes from the thin fluid film between droplet interface and
channel wall. Its reduced thickness leads to large velocity gradients and thus
also to viscous dissipation. As the droplet resistances at different volumes are
compared, it is not only the thin-film contribution which changes. Together with
the more elongated shape of the interfaces, also the precise shape of the
interior eddies change. The viscous dissipation within the droplet strongly
depends on the details of the eddies, in particular on the ratio between the
round backflows near the caps of the droplet and the rather ``Poiseuille-like''
parts in the interior of a long droplet. At this point, nothing can be said
about the relative importance of the viscous dissipation in the thin film and in
the eddies. We will return to this point below, in the discussion of the
distance between droplets.

In the experiment of Fig.~\ref{Ld_vs_vol} the range of droplet volumes is
limited, as large droplets tend to break or deform and wait in the incoming
junction of the loop: The waiting time in the junction causes an overestimation
of~$L_d$ by Eq.~\eqref{eq:small_loop}. Note that in the small-loop model of
Sec.~\ref{small} the new route is chosen instantaneously and does not take any
waiting time into account. There is also a lower bound for the applicable
droplet volume, given by the channel cross-section. As soon as the droplets have
no contact with the channel walls, their resistance strongly decreases. However,
we require their resistance to be large enough to overgrow the asymmetry of the
loop.

\subsection{Influence of the total flowrate}
\begin{figure}[tb]%
  \centering
  \includegraphics{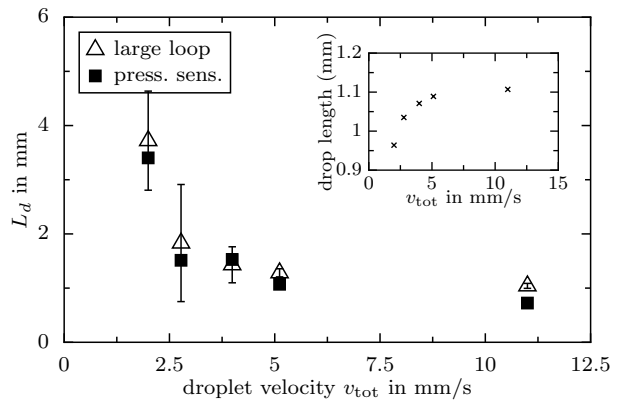}%
  \caption{The resistance length~$L_d$ (large-loop method and pressure sensor)
  as a function of the droplet velocity~$v_\tot$. Inset: length of the droplet
  (cap to cap). The parameters are~$\eta_d=0.01\,\pascal\,\second$,
  $\eta_\Phi=0.02\,\pascal\,\second$,
  $L_\short=53\,\milli\meter$,
  $L_\lon/L_\short=1.5$,
  $\Omega_d = 95\,\nano\liter$. The distance varies as 
  $\lambda$ from $5\,\milli\meter$ to $15\,\milli\meter$ from small to large
  velocities.}%
  \label{Ld_vs_v}
\end{figure}%
According to assumption~(C) the hydrodynamic resistance length~$L_d$ should
neither depend on the total flowrate in the channel, nor on the velocity of the
droplet. In fact, we found that this is the case only for velocities higher than
approximately $v_\tot=5\,\milli\meter/\second$, as Fig.~\ref{Ld_vs_v} indicates.
At small droplet velocities we find very large hydrodynamic resistances. The
droplet resistance decreases with increasing velocity, until only a weak
dependence is observed beyond $v_\tot=5\,\milli\meter/\second$. The data plotted
in Fig.~\ref{Ld_vs_v} have been obtained from the large-loop device and are in
good agreement with the reference measurement. It is important to stress this
agreement, as it shows that the large-loop device indeed measures the droplet
resistances, and that it does not depend on assumption~(C), as was
already explained at the end of Sec.~\ref{large}.

The velocity value~$v_\tot$ in the measurement is the one in the incoming channel, which
is systematically different from the true velocity at which the droplets move.
In the large-loop device, however, we find automatically balanced flowrates and
therefore also balanced velocities in the two branches, which allows to say that
both are half the incoming velocity, $v_\short\approx v_\lon\approx v_\tot/2$,
at least on average. That we obtain the true velocity immediately from a
measurement in the straight part of the channel is the reason why we performed
this experiment with the large-loop device and not with the small-loop device,
where the velocity ratios depend on the droplet resistances, unless the droplet
velocity is sufficiently high.

The measurements in Fig.~\ref{Ld_vs_v} have been performed using droplets of the
same volume and the same composition. The velocity was varied by injecting more
oil between the droplets. Together with the total flowrate, thus also the
initial spacing between the droplets was varied between $5\,\milli\meter$ and
$15\,\milli\meter$. The origin of the strong increase of resistance at small
velocities is unclear at the moment. We found a correlation with the shape of
the droplets (inset of Fig.~\ref{Ld_vs_v}), indicating a systematic lengthening
of the fast flowing droplets. This increases the thickness of the thin oil film
between the droplet and the walls and reduces the total resistance of the
droplets~\cite{WonRadMor95b}. A dynamical theory which would allow to fit the
droplet resistances given in Fig.~\ref{Ld_vs_v} is not known to us. In fact, we
tried to fit the data to existing formulae for the resistance as a function of
the droplet size and of the velocity: In Ref.~\cite{FuerstmanETAL07} a
proposition containing a scaling with $v^{1/3}$ is put forward for bubbles. In
Ref.~\cite{GrossETAL08} the scaling is rather~$v^{1/2}$. Neither formula yielded
a remotely satisfactorily result for the data in Fig.~\ref{Ld_vs_v}. The details
of the surrounding flow and of the internal circulation in the droplet are
evidently much more important than in the case of bubbles, where a simple
scaling approach seems to work~\cite{Bretherton61,FuerstmanETAL07,footnote}.

\subsection{Influence of the droplet viscosity}
\begin{figure}[tb]%
  \centering
  \includegraphics{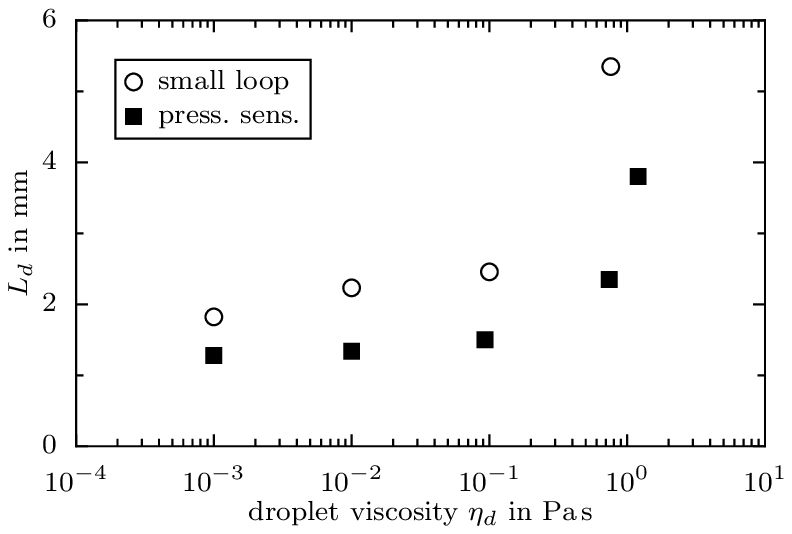}
  \caption{The resistance length~$L_d$ (small-loop method) as a function of the
  droplet viscosity.
  The parameters are~$\eta_\Phi=0.02\,\pascal\,\second$,
  $L_s=4.35\,\milli\meter$,
  $L_\lon/L_\short=1.18$,
  $v_\tot = 9\,\milli\meter/\second$,
  $\lambda > 9\,\milli\meter$, and
  $\Omega_d = 65\,\nano\liter$.}%
  \label{Ld_vs_eta}
\end{figure}%
Another set of experiments with the small-loop device explores the influence of
the viscosity of the glycerol/water mixture inside the droplets on their
resistance. The viscosity was changed by varying the concentrations of glycerol
and of water in the mixture, the continuous phase and the velocity were kept
constant. Figure~\ref{Ld_vs_eta} depicts the resulting resistance length~$L_d$
as a function of the fluid viscosity in the droplet. The values of~$L_d$ remain
approximately constant as long as the droplet viscosity is lower than the
viscosity of the surrounding phase. Beyond this value, they exhibit a strong
increase which can be attributed to the enhanced viscous dissipation by the
internal circulation in the droplets. Similar effects have been observed in
Ref.~\cite{GrossETAL08}. The agreement between the results of the small-loop
device and the pressure sensor is mediocre. We do not have a convincing
explanation for the deviations at the moment. Nevertheless, the two measurements
exhibit the same ascending trend. As described above, the velocity of the
droplets were quite fast, in order to stay within the part of Fig.~\ref{Ld_vs_v}
where the velocity does not strongly influence the resistance length. We chose
this regime in order to clearly separate the respective influence of the
viscosity from others.

The qualitative form of the flow fields are depicted in the sketch of the
counter-rotating eddies in Fig.~\ref{eddies}. The two panels of the figure
present two different topologies for the two extreme cases of the ratio
$\eta_d/\eta_\Phi$. These two topologies are expected because the fluid with
higher viscosity should avoid to produce several eddies, each of which increases
the viscous dissipation. Despite the fact that there are two clearly
distinguished regimes of the function $L_d(\eta_d)$ in Fig.~\ref{Ld_vs_eta}, in
one of which it is nearly constant, in the other fast growing, we find a rather
smooth change from one regime to the other, and not an abrupt one at the point
of equal viscosity~$\eta_d=\eta_\Phi$. The curve may further be extrapolated to
an infinite droplet viscosity, in which the result of a nearly filling rigid
object in a channel is reached. The resistance of such an object depends solely
on its filling property, that is on the thin film between object and walls. It
is then only the ratio of surface tension and exterior viscosity (exterior
Capillary number) which determines the thickness of the thin film.

\subsection{Influence of the initial inter-droplet distances}
\begin{figure}[tb]%
  \centering
  \includegraphics{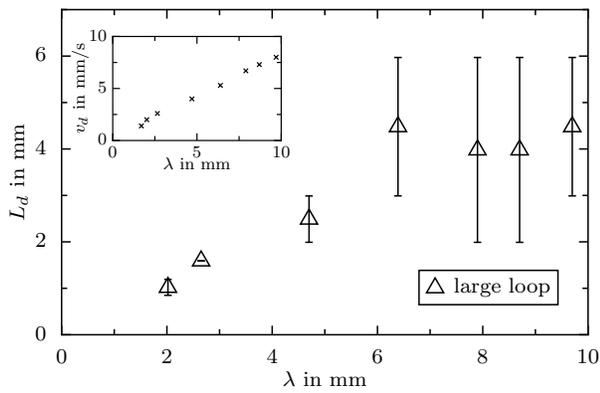}
  \caption{The resistance length~$L_d$ (large-loop method) as a function of the
  initial inter-droplet distance~$\lambda$. Inset: variation of the droplet
  velocity~$v_\tot$ with~$\lambda$.
  The parameters are~$\eta_d=1.4\,\pascal\,\second$,
  $\eta_\Phi=0.02\,\pascal\,\second$,
  $L_\short=36\,\milli\meter$,
  $L_\lon/L_\short=1.6$,
  $\Omega_d = 65\,\nano\liter$.}%
  \label{Ld_vs_lambda}
\end{figure}%
We now discuss possible cooperative effects between adjacent droplets and thus
explore the validity of assumption~(B). Of course, this can only be undertaken
by the large-loop device, since the small loop contains not more than one
droplet. Upon varying the initial distance between droplets, we find indeed a
critical droplet spacing below which assumption~(B) does not hold any more.
Figure~\ref{Ld_vs_lambda} visualizes this result, as measured by the large-loop
device. For small distances, an increase of the droplet resistance as a function
of the distance can be seen. Beyond approximately~$6\,\milli\meter$ we find an
apparent plateau. The large uncertainties in this part of the data are caused by
the small number of droplets in the branches of the loop. For any given maximal
size of the large loop results a maximal reasonable droplet distance. The
uncertainties can, however, be reduced either by using a ratio~$L_\lon/L_\short$
closer to unity, or by investigating narrower channels. In the latter case, the
droplet distance up to which cooperative effects occur, should be decreased
together with the width/height of the channel. The necessity to work at
distances larger than $6\,\milli\meter$ in order to guarantee the validity of
assumption~(B), has already been taken into account in the previously described
measurements.

The decrease of resistance at distances smaller than $6\,\milli\meter$ gives
evidence to a collective motion of the droplet. This means that a pair or series
of droplets can be displaced more easily than the same number of individual
droplets. This observation sheds light on the relative importance of the three
different contributions to the hydrodynamic resistance, as mentioned above in
the discussion of Fig.~\ref{eddies}. In the literature on (inviscid) bubbles,
the contribution from the thin film has been considered as the primary reason
for hydrodynamic resistance~\cite{Bretherton61,WonRadMor95b}. The decrease of
resistance observed in the measurements of Fig.~\ref{Ld_vs_lambda} proves that
in our case the thin-film contribution cannot be dominant but that the eddies
outside of the droplet contribute equally. From Fig.~\ref{Ld_vs_lambda} we can
estimate the size of the eddies. The incoming distances of $6\,\milli\meter$
correspond to a minimal distance between droplets of around~$3\,\milli\meter$ in
the branches where the droplet sequences are not regular anymore. The diameter
of the exterior eddies are then around $1\,\milli\meter$, thus two times the
width of the channel.

Note that for experimental reasons, the three relevant parameters, namely
droplet velocity, droplet spacing and the droplet volume cannot all be
independently controlled. Therefore, this set of experiments has not been
performed at constant flowrate (inset on Fig.~\ref{Ld_vs_lambda}). There is thus
a hidden influence of the droplet velocity on~$L_d$, which should pronounce the
dependence on~$\lambda$ even more.

\begin{figure}[tb]%
  \centering
  \includegraphics{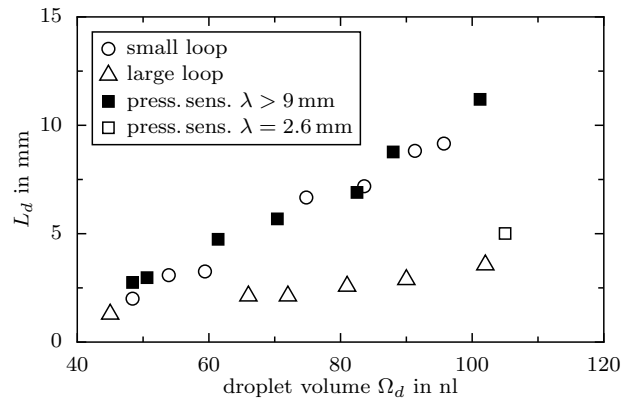}%
  \caption{Combined influence of two parameters: The resistance length~$L_d$
  (all three methods) as a function of the droplet volume. Plotted are the data
  from Fig.~\ref{Ld_vs_vol} (short-loop and pressure sensor), in comparison data
  using a smaller inter-droplet distance (pressure sensor and large loop,
  $\lambda\approx 3\,\milli\meter$). The empty symbol for the pressure sensor
  corresponds to a distance of $\lambda=2.6\,\milli\meter$. The other parameters
  are~$\eta_d=1.4\,\pascal\,\second$, $\eta_\Phi=0.02\,\pascal\,\second$,
  $v_\tot=6\,\milli\meter/\second$.}%
  \label{Ld_vs_vol2}
\end{figure}%

The full function $L_d(\lambda,v_\tot,\Omega_d,\eta_d,\eta_\Phi)$ can only be
fully displayed in a high-dimensional plot. A non-trivial dependence of the
parameters is found when varying two of them at the same time.
Figure~\ref{Ld_vs_vol2} shows such an example, namely the resistance
length~$L_d$ as a function of the drop volume. The small-loop data have already
presented in Fig.~\ref{Ld_vs_vol}. The corresponding large-loop resistances are
systematically smaller. The error bars and an independent measurement with the
pressure sensor both indicate that the difference is not an artifact. The
discrepancy is misleading, since the difference here is due to the smaller
inter-droplet distance of approximately $3\,\milli\meter$ in the large-loop
experiment. For the volume of~$65\,\nano\liter$, we find the same value
$L_d\approx2\,\milli\meter$ in Fig.~\ref{Ld_vs_lambda} at this very distance.
The resistance length, when measured with the small-loop device, is
approximately $5\,\milli\meter$ at a volume of $65\,\nano\liter$. This value is
approximately the one of the plateau at large distances in
Fig.~\ref{Ld_vs_lambda}. We may thus conclude that the small-loop device really
measures the resistances of individual droplets, and that this resistance
corresponds to the plateau at large distances in the large-loop device.

Figure~\ref{Ld_vs_vol2} leads us to a comment which we consider central in the
analysis of droplet traffic in networks. When droplets are driven through a
complex and possibly large network of microfluidic channels, comprising a lot of
junctions, then no direct control over parameters such as their distance or
their velocity is possible. When the droplet trains in an existing device
exhibit irregularities, i.e.~that they do not follow the trajectory they are
supposed to follow in order to fulfill the objective of the device, then it will
become necessary to understand the droplet dynamics in more detail. The present
work demonstrates some aspects of this understanding for a simple device, where
the mutual influence of the hydrodynamic resistance and the droplet movement
becomes apparent.

\section{Summary}

In the present paper we use three methods to measure the effective hydrodynamic
resistance of droplets in microchannels: One standard method based on the use of
pressure sensors and two variants of a new method relying on the analysis of
droplet trajectories in a simple microfluidic device. These two variants are
proposed as complementary techniques, as they are based on different model
assumptions on the droplet behavior in the microfluidic network.

The main idea put forward here is to use the global droplet dynamics, namely the
distances between droplets, their number in each branch of the loop, and the
choices taken at the first junction. All of them can simply be extracted by
analyzing a video of the droplets flowing. The methods are non-invasive as they
do not require additional measuring channels for pressure sensors, nor do they
require expensive equipment. The interpretation of these observables to obtain
the resistances is based on a theoretical model, the simplifying assumptions of
which we test in the present work. The proposed methods focus rather on the
understanding of the droplet dynamics than on the isolated measurement of the
all-parameter dependence of the hydrodynamic resistance of droplets. In this
sense, it is not meant to replace the standard method employing pressure
sensors, but rather to provide a guided example for how the droplet dynamics
reacts on variations of the hydrodynamic resistance and vice versa.

We were able to show that there exists a parameter regime where the assumptions
adopted by the model do hold. In this regime, the quantity that we extracted
from the experimental observations indeed serves as the wanted hydrodynamic
resistance of droplets. These results are presented in Figs.~\ref{Ld_vs_vol},
\ref{Ld_vs_v}, \ref{Ld_vs_eta}, \ref{Ld_vs_lambda}, and~\ref{Ld_vs_vol2}, where
we varied different parameters, namely the volume, velocity and viscosity of
droplets, as well as their distance. All these sets of measurements were done in
a way such that they avoid those parameter regimes where the model assumptions
are not valid. Therefore, they do not contradict each other. This claim is
enforced by independent and more direct measures of the droplet resistances,
undertaken with an expensive and elaborate pressure-sensor technique. The
requirement to find the (formerly unknown) valid part of a high-dimensional
parameter space presented the main intellectual difficulty while generating the
results displayed in the figures. The parameter dependencies we found are the
following: The hydrodynamic resistance increases with increasing volume and
increasing viscosity of the droplet. It increases with inter-droplet distances
as long as the distances are smaller than approx. $6\,\milli\meter$. Beyond that
value the resistance does not depend on the inter-droplet distance. For small
droplet velocities, the resistance strongly decreases with increasing velocity.
It changes only little for velocities beyond $3\,\milli\meter/\second$. On a
more fundamental level of description, the results presented here demand a
systematic determination of the hydrodynamic resistance as a function of droplet
volume, spacing, velocity and viscosity. This could be done for example by
extending Refs.~\cite{LinkETAL04,AnnBonSto03}, which are concerned with the
resistance of inviscid bubbles, not droplets. In the limit of high velocities,
large distances, and small volumes, the resistance will still have to turn out
to be constant, as assumed by the simple model used here.

\acknowledgments{%
This work was supported by the French National Agency for
Reseach (ANR No\,06--NANO--048--02) and the National Centre de la Recherche
Scientifique~(CNRS). We thank all the participants in this project for
stimulating discussions.
}



\begin{thebibliography}{99}

\bibitem{JoaAjd05}
  M. Joanicot and A. Ajdari,
  Science~\textbf{309}, 887 (2005).

\bibitem{CristobalETAL06}
  G. Cristobal, L. Arbouet, F. Sarrazin, D. Talaga, J.-L. Brunel, M. Joanicot and L. Servant,
  Lab on a Chip~\textbf{6}, 1140 (2006).

\bibitem{TiceETAL03}
  D. Tice, H. Song, A. D. Lyon and R. F. Ismagilov,
  Langmuir~\textbf{19}, 9127 (2003).

\bibitem{DenTsoHatDoy05}
  D. Dendukuri, K. Tsoi, T. A. Hatton and P. S. Doyle,
  Langmuir~\textbf{21}, 2113 (2005).

\bibitem{ZheRoaIsm03}
  B. Zheng, L.S. Roach and R. F. Ismagilov,
  J. Am. Chem. Soc.~\textbf{125}, 11170 (2003).

\bibitem{PraGer07}
  M. Prakash and N. Gershenfeld,
  Science~\textbf{315}, 832 (2007).

\bibitem{ThorsenETAL01}
  T. Thorsen, R. W. Roberts, F. H. Arnold, and S. R. Quake,
  Phys. Rev. Lett.~\textbf{86}, 4163 (2001).

\bibitem{SonTicIsm03}
  H. Song, J.D. Tice, and R. F. Ismagilov,
  Angew. Chem., Int. Ed.~\textbf{42}, 767 (2003).

\bibitem{WilBarKloMaiTab06}
  H. Willaime, V. Barbier, L. Kloul, S. Maine, and P. Tabeling,
  Phys. Rev. Lett.~\textbf{96}, 054501 (2006).

\bibitem{GarFueWhi05}
  P. Garstecki, M.J. Fuerstman, and G.M. Whitesides,
  Nature Phys.~\textbf{1}, 168 (2005).

\bibitem{SchAjd08}
  M. Schindler and A. Ajdari,
  Phys. Rev. Lett.~\textbf{100}, 044501 (2008).

\bibitem{AdzVel06}
  B. J. Adzima and S. S. Velankar,
  J. Micromech. Microeng.~\textbf{16}, 1504 (2006).

\bibitem{MorOkkBru05}
  N. A. Mortensen, F. Okkels, H. Bruus,
  Phys. Rev. E~\textbf{71}, 057301 (2005).

\bibitem{EnglETAL05}
  W. Engl, M. Roche, A. Colin, P. Panizza and A. Ajdari,
  Phys. Rev. Lett.~\textbf{95}, 208304 (2005).

\bibitem{Bretherton61}
  F. P. Bretherton,
  J. Fluid Mech.~\textbf{10}, 166 (1961).

\bibitem{HodJenRal04}
  S. R. Hodges, O. E. Jensen, and J. M. Rallison,
  J. Fluid Mech.~\textbf{501}, 279 (2004).

\bibitem{WonRadMor95b}
  H. Wong, C. J. Radke, and S. Morris,
  J. Fluid Mech.~\textbf{292}, 95 (1995).

\bibitem{LinkETAL04}
  D. R. Link, S. L. Anna, D. A. Weitz, H. A. Stone,
  Phys. Rev. Lett.~\textbf{92}, 054503 (2004).

\bibitem{FuerstmanETAL07}
  M. J. Fuerstman, A. Lai, M. E. Thurlow, S. S. Shevkoplyas, H. A. Stone, and G. M. Whitesides,
  Lab Chip~\textbf{7}, 1479 (2007).

\bibitem{QinXiaWhi96}
  D. Qin, Y. Xia and G. M. Whitesides,
  Adv. Mater.~\textbf{8}, 917 (1996).

\bibitem{DufMcDSchWhi98}
  D. C. Duffy, J. C. McDonald, O. J. A. Schueller and G. M. Whitesides,
  Anal. Chem.~\textbf{70}, 4974 (1998).

\bibitem{HarrisonETAL04}
  C. Harrison, J.T. Cabral, C.M. Stafford, A. Karim and E. J. Amis,
  J. Micromech.~Microeng.\textbf{14}, 153 (2004).

\bibitem{MalschETAL08}
  D. Malsch,  M. Kielpinski, R. Merthan, J. Albvert, G. Mayer, J. M. K\"ohler, H. Sube, M. Stahl, T. Henkel,
  Chem. Eng. J.~\textbf{135}, S166 (2008).

\bibitem{GrossETAL08}
  G. A. Gro\ss, V. Thyagarajan, M. Kielpinski, T. Henkel, J. M. K\"ohler,
  Microfluid Nanofluid~\textbf{5}, 281 (2008).

\bibitem{footnote}
  It has to be noted that the large deviations in the data fit in
  Ref.~\cite{FuerstmanETAL07} put into question the scaling argument first put
  forward by \cite{Bretherton61}. We do not have a better explanation,
  however.

\bibitem{AnnBonSto03}
  S. L. Anna, N. Bontoux, and H. A. Stone,
  Appl. Phys. Lett.~\textbf{82}, 364 (2003).

\end{thebibliography}
\end{document}